\documentclass[onecolumn,aps,preprint,showpacs,amsmath,amssymb]{revtex4-1}
\usepackage{graphicx}
\usepackage{latexsym}
\usepackage{dcolumn} 
\usepackage{color}
 
\usepackage{bm}

\newcommand{\parenth}[1]{\left( #1 \right)}
\newcommand{\bparenth}[1]{\left[ #1 \right]}

\begin{document} 

%\documentclass[pra,preprint,tightenlines,showpacs,array,amscd,asmath, 
%hhline,epsf,amssymb]{revtex4}
%\usepackage{graphicx}
%\usepackage{graphics}
%%%%%%%%%%%%%%%%%%%%%%%%%%%%%%%%%%%%%%%%%%%%%%%%%%%%%%%%%%%%%%%%%%

%\begin{document}
\newcommand{\eq}{\begin{equation}}
\newcommand{\eqe}{\end{equation}}

\title{Self-similarity analysis of the non-linear Schr\"odinger equation in the Madelung form}

\author{Imre F. Barna$^{1,2}$ and Mih\'aly A. Pocsai$^{1,3}$ and L. M\'aty\'as$^{4}$}
\address{ $^1$ Wigner Research Center for Physics of the Hungarian Academy of Sciences 
\\ Konkoly-Thege Mikl\'os \'ut 29 - 33, 1121 Budapest, Hungary \\
$^2$  ELI-HU Nonprofit Kft.,  Dugonics T\'er 13, H-6720 Szeged, Hungary \\
$^3$ University of P\'ecs, Institute of Physics, Ifj\'us\'ag \'utja 6 H-7624 P\'ecs, Hungary \\ 
$^4$Sapientia University, Department of Bioengineering, Libert\u{a}tii sq. 1, 530104 Miercurea Ciuc, Romania  }
\date{\today}
%%%%%%%%%%%%%%%%%%%%%%%%%%%%%%%%%%%%%%%%%%%%%%%%%%%%%%%%%%%%%%%%%%%
\begin{abstract}  
In the present study a particular case of Gross-Pitaevskii or non-linear Schr\"odinger 
equation is rewritten to a form similar to a hydrodynamic Euler 
equation using the Madelung transformation. 
The obtained system of differential equations is highly nonlinear. 
Regarding the solutions, a larger coefficient of the nonlinear term yields 
stronger deviation of the solution from the linear case. 
\end{abstract}

%\pacs{34.10.+x, 34.50.-s, 34.50.Fa}

\maketitle

%%%%%%%%%%%%%%%%%%%%%%%%%%%%%%%%%%%%%%%%%%%%%%%%%%%%%%%%%%%%%%%%%%%%%%%%%%%
\section{Introduction}
 
In 1926, at the advent of the quantum mechanics Madelung \cite{madelung1, madelung2} gave a hydrodynamical foundation of the Schr\"odinger equation and raised interpretation questions which are open till today. Later, people realized that the Madelung Anzatz is just the complex Cole-Hopf transformation \cite{hopf,cole} which is sometimes applied to linearize non-linear partial differential equations(PDEs). The Madelug form of the Schr\"odiger equation has a remarkable  property,  that Planck's constant  (representing quantum mechanics) appears only once, namely as the coefficient of the quantum potential. This quantity also appears in the non-mainstream attempt of quantum mechanics  (the de Broglie-Bohm pilot wave theory \cite{bohm1} (in other context))  which interprets quantum mechanics as a deterministic non-local theory. The Madelung transformation may create a common language between scientist of fluid dynamics and quantum mechanics and can be fruitful to give an additional interpretation of quantum mechanics. In the last decades, hydrodynamical description of quantum mechanical systems become a popular technical tool in numerical simulations. Detailed reviews in this field can be found in a booklet \cite{qtraj} or in Wyatt's monograph \cite{wyatt}. 

The application of the original Madelung (or to say Cole-Hopf) transformation grew out the linear quantum theory and were applied to more sophisticated field theoretical problems. Recently, Chavanis and Matos \cite{curved} applied the Madelung Ansatz to the Klein-Gordon-Maxwell-Einstein equation in curved space-time possibly describing self-gravitating Bose-Einstein condensates, coupled to an electromagnetic field. This study clearly presents the wide applicability of the original Madelung Ansatz to refined quantum field theories with the possible hydrodynamic interpretation picture.  

%Some years ago Tsubota {\it{et al.}} \cite{mak} investigated the superfluid helium and the Bose-Einstein condensate as systems which can be 
%described with the non-linear Schr\"odinger(NLS) equation transformed with the Madelung Ansatz to equations which are called quantum hydrodynamics. This review gives the starting point of our forthcoming analysis.  
%It is well-known that the NLS equation has soliton solutions which are special kind of traveling waves without dispersion \cite{ablow}. (The %corresponding literature become enormous, which we skip now.)  It is also clear, that NLS has no analytic dispersive solutions therefore our 
%self-similar Ansatz \cite{sedov} leads to contradiction.    

As an analytical tool of investigation we use the self-similar Ansatz of ref. \cite{sedov}.
It is a powerful method to study the global properties of the dispersive solutions of various non-linear partial-differential equations (PDEs)
In other words, the intermediate asymptotic of the problem is described. Therefore this solution is valid when the precise initial conditions are no longer important, however, the final steady state is not yet reached. For real physical systems this region attracts the most attention. 
In contrast to the full solution it is much simpler and easier to understand for the entire parameter space. 
The second reason for studying them is that they are solutions of a system of ordinary differential equations(ODEs)  (not a PDE) and thus do not suffer the full inherent numerical problems of the full PDEs.  Most of the cases this tool helps to understand some of the global properties of the 
solution eg. shock waves or the existence of compact supports. First, we applied this Ansatz to the modified Cattaneo-Venotte heat-conduction problem \cite{imre_robi} and obtained a compact solution with finite propagation speed. Later, we successfully generalized the Ansatz to multi space dimensions and studied numerous viscous fluid equations \cite{imre0,imre1} ending up with a book chapter of \cite{imre_book}.  As main results we are now able to interpret the birth of Benard cells in heated fluids \cite{imre2}.    
 
In our former study \cite{imre_mad} we investigated the Schr\"odinger equation in the Madelung form with the self-similar Ansatz and presented analytic solutions with discussion. Unlike to classical fluids the obtained density of the Madelung fluid has infinite number of zeros which 
is a fingerprint of the quantum mechanical origin.   
 Now we analyze the Madelung form of non-linear Schr\"odinger(NLS) equation with the same Ansatz, and compare 
the obtained results to the previous linear one.  As far as we know,  there are no direct clear-cut analytic solutions existing for the Madelung form of the NLS equation by now. It is important to mention that celebrated standard numerical algorithms exist for solving the NLS \cite{num1,num2}. We hope that our present study helps to attract attention to this question. Parallel to the well-known quantum mechanical picture there is a looming hydrodynamical world hidden behind the Schr\"odinger equation from the very beginning which is poorly understood. 

%%%%%%%%%%%%%%%%%%%%%%%%%%%%%%%%%%%%%%%%%%%%%%%%%%%%%%%%%%%%%%%%%%%%%%%%%%%
\section{Theory and results} 

In a weakly interacting Bose system, the macroscopic wave function $\Psi$ appears as the order parameter of the Bose-Einstein condensation, 
obeying the Gross-Pitaevskii(GP) equation  \cite{Gr61,Pi61,DaGiPiSt99,ErScYa10}
\eq
i\hbar \frac{\partial \Psi}{\partial t } = \left( -\frac{\hbar^2 \nabla^2}{2m} + n |\Psi|^2 - \mu \right)\Psi.  
\eqe 
Here, $n = 4\pi \hbar^2 a /m$ represents the coupling constant characterized by the s-wave scattering length a.  
The particle mass and the chemical potential are $m$ and $\mu$, respectively.  
Due to the second (non-linear) term of the right hand side of this equation, it is definitely different from the linear Schr\"odinger equation \cite{imre_mad}, therefore the following analysis with the presented results are independent as well.   

The role and the properties of the GP system has been investigated from various viewpoints.  Without completeness we mention some of them. 
The question of collective excitations of the trapped Bose gases was analyzed by \cite{csord}. 
Later, it is considered in the study of self-organization of the Bose-Einstein(BE) condensates \cite{doma}. 
  
Certain generalized versions of Eq. (1) have been also applied to the study of BE condensate of polaritons 
\cite{WoCa2007, LaWo2008, Roumpos2011, OsAb2012,RoKe2014, BaMaMi2015,LiEgMa2015,BeKe2017, ScWiOs2017}.

Considering the Madelung Ansatz for the order parameter with the amplitude and phase $\Psi = \sqrt{\rho}e^{i\phi}$ 
where the density of the condensate is $\rho$ and the superfluid velocity is ${\bf{v}} = (\hbar/m)\nabla \phi$. 
(The expression "superfluid velocity" is taken from Tsubota {\it{et al.}} \cite{mak} and means - in this content - that the fluid in inviscid.)

The obtained hydrodynamical equations - the continuity and the Euler - are the following, 
\begin{eqnarray}
\frac{\partial \rho}{\partial t} + \nabla \cdot (\rho {\bf{v}}) &=& 0, \nonumber \\
\frac{\partial {\bf{v}}}{\partial t} + \bf{v}   \cdot \nabla  {\bf{v}} &=&     \frac{\hbar^2}{2 m^2} 
\nabla\left(  \frac{\triangle \sqrt{\rho}}{\sqrt{\rho}} \right)  -\frac{\nabla}{m\rho} \left(  \frac{n \rho^2}{2} \right),   
\label{mad_euler}
 \end{eqnarray}
where we consider free motion (no extra potential energy term U({\bf{r}}) is used) and the chemical potential $\mu$ was set to zero as well.  It is easy to show that an additional chemical potential term $\mu$ gives no contribution to the final hydrodynamical equations. The role of the external potential energy term $U({\bf{r}})$ will be mentioned later. 

The first term of the right hand side of the second equation is the quantum pressure term. The second term makes the difference to the original Madelung equations, this is a kind of {\it{second quantum pressure}} due to the non-linear origin of the GP equation.     
Note, that these are most general vectorial equation for the velocity field ${\bf{v}}$ which means that one, two or three dimensional motions 
can be investigated as well. In the following we will consider the two dimensional flow motion  ${\bf{v}} = (u,v) $ in Cartesian coordinates. 
The functional form of the three and one dimensional equation system will be mentioned briefly, too.  

We search the solution of Eq.(\ref{mad_euler}) in the form of the self-similar Ansatz \cite{sedov},  	
\eq 
\rho(x,y,t)=t^{-\alpha}f\left(\frac{x+y}{t^\beta}\right):=t^{-\alpha}f(\eta),
\hspace*{5mm}  u(x,y,t)=t^{-\delta}g(\eta),  \hspace*{5mm} 
v(x,y,t)=t^{-\epsilon}h(\eta), 
\label{self}
\eqe 
where $f,g$ and $ h$ are the shape functions of the density and the velocity field, respectively. 
The similarity exponents $\alpha, \beta, \delta, \epsilon $ are of primary physical importance since $\alpha, \delta, \epsilon $  
represent the damping of the magnitude of the shape function. The exponent $\beta$ represents the
spreading of the solution. Additional technical details about the general features of the Ansatz can be found in 
our former papers \cite{imre1,imre2}.  Except some extraordinary cases only  
all positive similarity exponents correspond to physically relevant dispersive solutions which mean decaying features at $x,y,t  \rightarrow \infty $. 
Usually negative exponents describe exploding solutions. Substituting the Ansatz (\ref{self}) into (\ref{mad_euler}) 
and executing additional algebraic manipulation the following ODE system can written down for the shape functions
\begin{eqnarray}
-\alpha f - \frac{\alpha}{2}f'\eta + f'g + fg' + f'h + fh' &=& 0, \nonumber \\ 
-\left( \frac{3}{2} \alpha -1\right)g - \frac{\alpha}{2}g'\eta + gg' + hg' &=& 
\frac{\hbar^2}{2m^2}\left(\frac{f'^3}{2f^3} -  \frac{f'f''}{f^2} +
  \frac{f'''}{2f}\right) - \frac{n}{m}f',  \nonumber \\ 
 -\left( \frac{3}{2} \alpha -1\right)h - \frac{\alpha}{2}h'\eta  + gh' + hh' &=&  
 \frac{\hbar^2}{2m^2}\left(\frac{f'^3}{2f^3} -  \frac{f'f''}{f^2} +
  \frac{f'''}{2f}\right) - \frac{n}{m}f',
\label{syst}
\end{eqnarray}
where prime means derivation with respect to the new variable $\eta$. 
The relations among the similarity exponents are the following:  $\alpha $ can be taken arbitrary, $\beta = \alpha/2$ and 
$\epsilon = \delta = 3\beta-1 = 3\alpha/2 -1$. 
At this point we have to emphasize, that a particular form of the external potential 
$U(x,y,t)$ can be added to (1) without leading to contradiction among the similarity exponents. The form in 2D is the following: $U(x,y,t) = \frac{1}{t^{2\beta}} \cdot  v(\eta)$ where $v(\eta)$ can be any kind of function with existing first derivative. 

(Of course, there is no evidence that the obtained modified (\ref{syst}) could be solved with analytical means.) 

Note, that the ODE system of the linear Schr\"odinger equation \cite{imre_mad}  has a much simpler form of 
\begin{eqnarray}
-\frac{1}{2}f - \frac{1}{2}f'\eta + f'g + fg' + f'h + fh' &=& 0, \nonumber \\ 
-\frac{1}{2}g - \frac{1}{2}g'\eta + gg' + hg' &=& 
\frac{\hbar^2}{2m^2}\left(\frac{f'^3}{2f^3} -  \frac{f'f''}{f^2} +
  \frac{f'''}{2f}\right),  \nonumber \\ 
-\frac{1}{2}h - \frac{1}{2}h'\eta  + gh' + hh' &=&  
 \frac{\hbar^2}{2m^2}\left(\frac{f'^3}{2f^3} -  \frac{f'f''}{f^2} +
  \frac{f'''}{2f}\right),
\label{syst1}
\end{eqnarray}
where all the similarity exponents have the fixed value of $+1/2$. Our experience tells that this is
usual for regular heat conduction, diffusion or for non-compressible Navier-Stokes
equations \cite{imre_book}.
For the linear case after some additional trivial algebraic steps 
a decoupled ODE can be derived for the shape function of the density 
\eq
2f''f - (f')^2 + \frac{m^2 \eta^2f^2}{2\hbar^2} = 0. 
\label{dens}
\eqe  
 
For one, two or three dimensions the multiplicative factor of $\hbar$ in the last term is different. It is 1, 2 or 3, respectively. 
%%%%%%%%%%%%%%%%%%%%%%%%%%%%%%%%%%%%%%%%%%%%%%%%%%%%%%%%%%%%%%%%%%%%%%%%%%%%%
%\begin{figure}
%	\includegraphics[width=0.9\textwidth]{Fig1.pdf}
%	\caption{The solution of Eq.~(\ref{bess})  ($c_1=c_2=1$) the yellow curve is for $m =1$ and blue curve is for $m= 0.5$.}
%\label{fig:f_m1_m2}     
%\end{figure}
%%%%%%%%%%%%%%%%%%%%%%%%%%%%%%%%%%%%%%%%%%%%%%%%%%%%%%%%%%%%%%%%%%%%%%%%%%%%%
%\begin{figure}
%	\includegraphics[width=0.9\textwidth]{f_eta_c1_c2.pdf}
%	\caption{The solution of Eq.~(\ref{bess})  ($m=1$) the blue curve is for $c_{1} = c_{2}  = 1$, the yellow for $c_{1} = 1/4$ and $c_{2} = 1/2$ and the green for $c_{1} = -1/2$ and $c_{2} = 1/2$.}
%\label{fig:_c1_c2}     
%\end{figure}
%%%%%%%%%%%%%%%%%%%%%%%%%%%%%%%%%%%%%%% 
 
With the help of Bessel functions \cite{NIST} the solution of (\ref{dens}) can be expressed 
\begin{equation}
	f \parenth{\eta} = \frac{\pi \eta}{64} \bparenth{c_{1} J_{1/4} \parenth{\frac{\sqrt{2} m \eta^{2}}{8}} - c_{2} Y_{1/4} \parenth{\frac{\sqrt{2} m \eta^{2}}{8}}}^{2}, 
\label{solution1}
\end{equation}
where $c_1$ and $c_2$ are the usual integration constants. 
All the additional properties of this solution can be found in our former paper \cite{imre_mad}.  
(There is a misprint in the final \cite{imre_mad} at Eq. 7, the parenthesis is missing in front of the exponent 2. This is the correct form.) 

 The most general equation of  Eq. (\ref{syst}) cannot be fully integrated by analytical means. 
The first ODE of such systems \cite{imre_book} explicitly means mass conservation. Furthermore in the most cases it can be integrated 
analytically. However, the first equation of  (\ref{syst}) is not a total derivative 
for any $\alpha$. This is highly unusual among our investigated systems till now.  
As a second case let's consider $(3/2\alpha -1) = \alpha/2$ this gives $\alpha = 1, \beta = \delta = \epsilon = 1/2$.  Now the sum of the second and third equation of 
(\ref{syst1}) can be integrated leading to
\eq
\frac{g^2}{2} + g\left(h-\frac{\eta}{2} \right) + \frac{h^2}{2}  +\frac{h\eta}{2} + \frac{2nf}{m} - \frac{\hbar^2}{2m^2} \left( -\frac{f'^2}{4f^2} + 
\frac{f''}{2f} \right)+ c_1 = 0. 
\label{exper}
\eqe 
For the sake of simplicity we set $c_1$ to zero. 
Note, that the variables $ g,f $ and $h$ are still coupled in this single equation. 
However, (\ref{exper}) is quadratic in $g(\eta)$ and the solutions  can be expressed with the well-know formula of  
\eq
g_{1,2} = \frac{\eta}{2} - h \pm \sqrt{  
 \frac{\eta^2}{4} - \frac{4nf}{m}  + \frac{\hbar^2}{2m^2}\left( -\frac{f'^2}{4f^2}  + \frac{f''}{2f} \right)   }.
\label{second}
\eqe
Even this includes both functions of $h$ and $f$. To avoid this problem let's fix the discriminant to zero, which means 
that the velocity shape function will be single valued.  
The remaining ODE reads:  
\eq
2 ff'' - (f')^2 + f^2\left( \frac{\eta^2 m^2}{2 \hbar^2} - \frac{8nmf}{\hbar}\right) = 0
\label{dens2}
\eqe
 with the additional restraint that $4f^2 \ne 0$ meaning that the physical density of a fluid must be positive. 
Note, that without the last term we get back Eq. (\ref{dens}) with the solution of Eq. (\ref{solution1}). 
We tried numerous variable transformations unfortunately, we could not find any closed solution for the full Eq. (\ref{dens2}). 
(Note, that if the coefficient of the second term would be twice as the first term there would be a possible solution for the full equation including
the third order term.) 
It is also clear from Eq. (\ref{second}) that the velocity shape functions have the trivial form of $ g + h = \eta/2 = \frac{x+y}{2\sqrt{t}}$ 
which is the same as for the linear case.  

\begin{figure}
\includegraphics[width=0.9\textwidth]{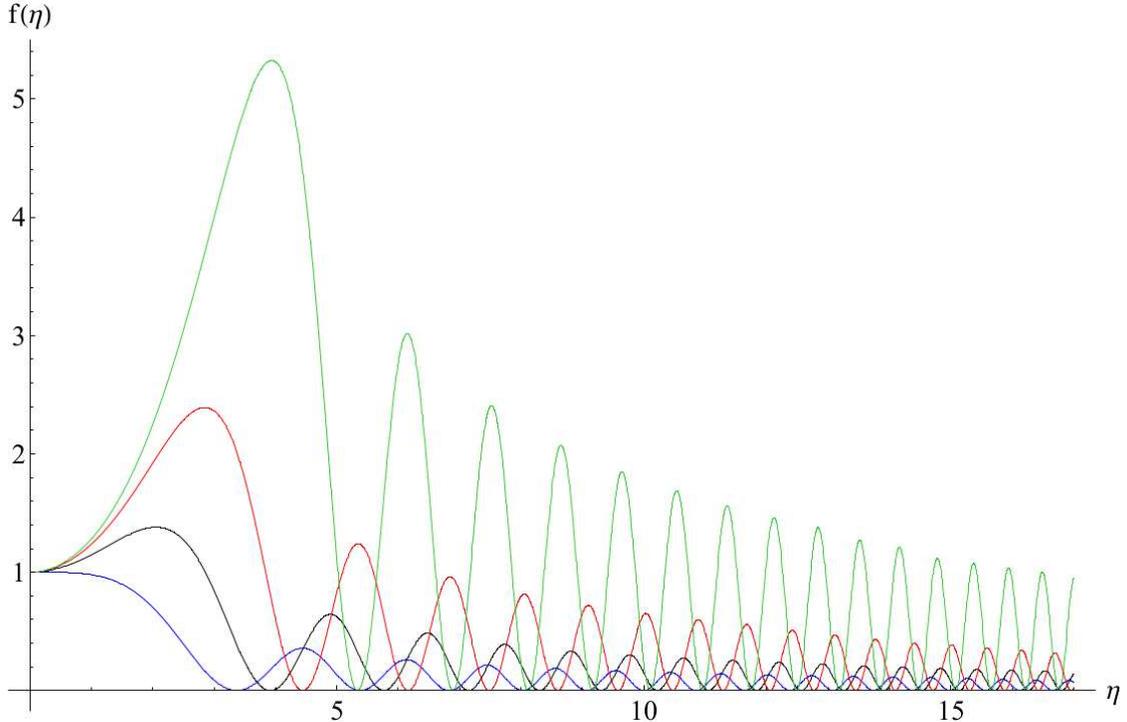}
	\caption{Various shape functions of the density. The blue curve is the analytic solution of Eq. (\ref{solution1}) with $n=0$ nonlinearity parameter, 
Black, red and green function are numerical results of Eq. (\ref{dens2}) for $n= 0.078,  \> 0.11$ and $ 0.13$ nonlinearity parameters. 
(All curves are for the same initial conditions of $f(0) = 1$  and $f'(0) = 0$.)
Note, that larger parameters mean larger deviation from the analytic result. 
  }
\label{f_eta}
\end{figure}

Figure \ref{f_eta}  shows the analytical solution of Eq. (\ref{solution1}) and numerically solution of Eq. (\ref{dens2}) for various nonlinear parameters 
$n$ with the same initial conditions of $f(0) = 1 $ and   $f'(0) = 0 $. The $\hbar  = m = 1$ units are used.  
All solutions have a strong damping with stronger and stronger oscillations at large arguments. 
The functions are positive for all values of the argument, (which is physical for a 
fluid density), but such oscillatory profiles are completely unknown in regular fluid mechanics \cite{imre_book}. The most interesting feature is the infinite number of zero values which cannot be interpreted physically for a classical real fluid. Remembering the restraint $4f^2 \ne 0$ of the calculation 
the solution function falls into infinite number of finite intervals.  We think that this is a clear fingerprint that the obtained 
Euler equations have a quantum mechanical origin.   

The major difference to the analytic solution 
is the following the $\int_0^{\infty}  f(\rho) d\rho$ can be evaluated (now of course just numerically up to an arbitrary $\rho_{max}$) 
giving us a convergent integral. This clearly shows that the wave function of the corresponding quantum mechanical system (see below) is the element of the $L^2$ space. We investigated numerous initial conditions all giving the same result of a finite integral. Of course, this is just an empirical statement and not a rigorous mathematical proof.  
  
The $\sqrt{\rho}$ function could be interpreted as the fluid mechanical analogue of the real part of the wave function of the free GP quantum mechanical particle.  For the linear case the square root of  Eq. (\ref{solution1})  is a kind of far analogue of the original free Gaussian wave packet. Unfortunately, we cannot find any direct transformation between these two functions.  

To obtain the complete original wave function, the imaginary part has to be evaluated as well. It is simple from 
$\eta = \frac{x+y}{t^{1/2}}= 2(g+h)$ that 
\eq
 S = \frac{m}{\hbar}\int_{\bf{r}_0}^{\bf{r}_1}  {\bf{v}}{\bf{dr}} = 
\frac{m}{\hbar}\frac{(x+y)^2}{4t}. 
\eqe 

\begin{figure}
\includegraphics[width=0.9\textwidth]{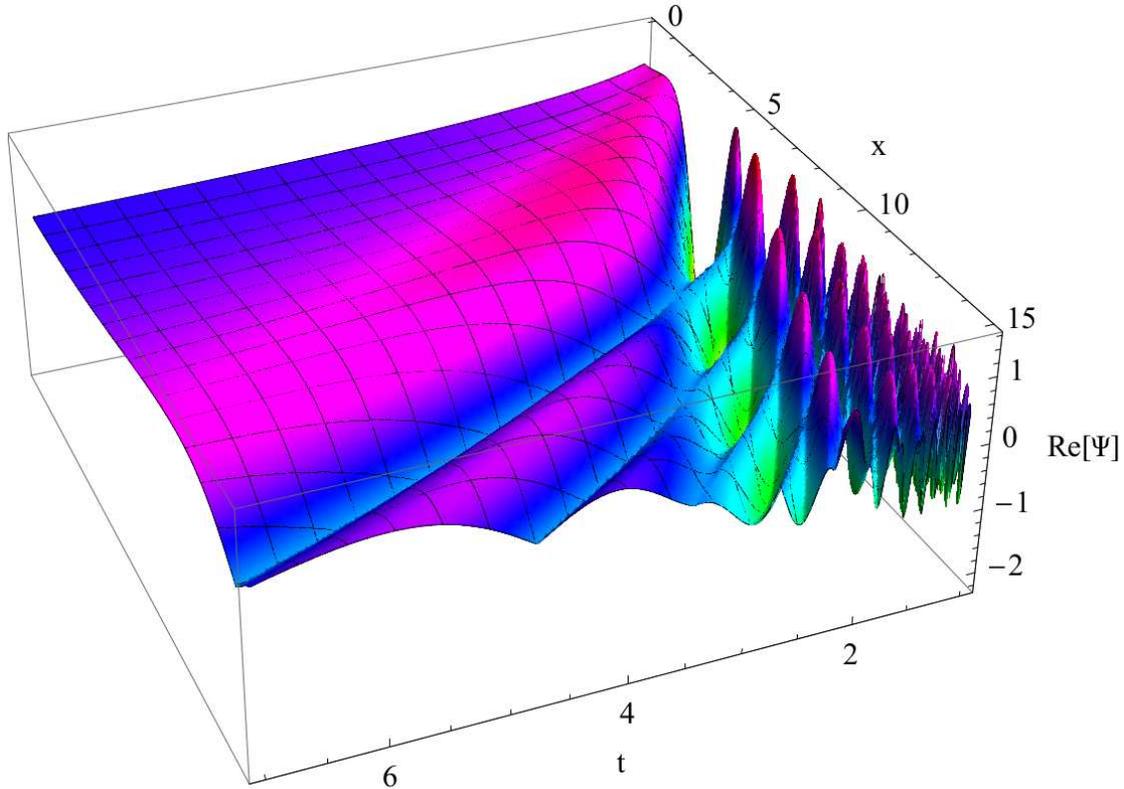}
	\caption{The projection of the real part of the wave function (\ref{wavf}) along the x-t plane.  }
\label{wavefunc}
\end{figure}

Now
\begin{equation} 
	\Psi(x,y,t) =\sqrt{ \rho(x,y,t)}e^{iS(x,y,t)}  =  \sqrt{ t^{-1} \cdot f \left( \frac{[x+y]}{t^{1/2}}  \right) } e^{\frac{i m}{\hbar} \frac{(x+y)^{2}}{4t,}}
\label{wavf}
\end{equation}
where $f(\eta)$ is now a numerical function. 
Figure \ref{wavefunc} shows the projection of the real part of the wave function to the ${x,t}$ sub-space.  
At small times there are clear oscillations, but at larger times the strong damping is evident.  
If the time is fixed ($ t = t_0 $ so to say) there are infinite number of spatial oscillations which can be interpreted as 
a quantum mechanical heritage of the system. 

%%%%%%%%%%%%%%%%%%%%%%%%%%%%%%%%%%%%%%%%%%%%%%%%%
%%%%%%%%%%%%%%%%%%%%%%%%%%%%%%%%%%%%%%%%%%%%%%%%%%%%%%%%%%
%%%%%%%%%%%%%%s%%%%%%%%%%%%%%%%%%%%%%%

%quadratic argument of the Bessel functions the power spectra has a strong decay as well.  
%%%%%%%%%%%%%%%%%%%%%%%%%%%%%%%%%%%%%%%%%%%%%%%%%
%%%%%%%%%%%%%%%%%%%%%%%%%%%%%%%%%%%%%%%%%%%%%%%%%%%%%%%%%%
%%%%%%%%%%%%%%%%%%%%%%%%%%%%%%%%%%%%
 \section{Summary and Outlook}
We gave a short historical review of the interpretation and development of the Madelung equation.   
As second point we introduced (the not so well known) self-similar Ansatz which is a robust tool to investigate the local and global properties 
of of non-linear PDEs. Finally, the Madelung form of the free NLS equation was investigated and the results were compared to the ones obtained from the linear Schr\"odinger equation. We think that the highly oscillatory fluid density (with the infinite number of zeros) are a clear fingerprint of the quantum mechanical origin of the system. We also believe that the noticeable quadratic argument of the solutions might get an interpretation in the future.  
However, the presented approach is still evolving and works are in progress. We plan to apply our method to the Madelung equations of more general 
non-linear Schr\"odinger equations like \cite{remi}.      
%%%%%%%%%%%%%%%%%%%%%%%%%%%%%%%%%%%%%%%%%%%%%%%%%%%%%%%%%%%%%%% 
%
\section{Acknowledgment} 
The ELI-ALPS project (GINOP-2.3.6-15-2015-00001) is supported by the European Union and co-financed by the European
Regional Development Fund. 
%We thank for Prof. Andr\'as Csord\'as for useful comments.  
%
%%%%%%%%%%%%%%%%%%%%%%%%%%%%%%%%%%%%%%%%%%%%%%%%%%%%%%%%%%%%%%%%%%%

 %%%%%%%%%%%%%%%%%%%%%%%%%%%%%%%%%%%%%%%%%%%%%%%%%%%%%%
\end{document}